\documentclass[%
 reprint,
amsmath,amssymb,
aps,
prb,
]{revtex4-2}

\usepackage{graphicx}
\usepackage{dcolumn}
\usepackage{bm}
\usepackage{ulem} 
\usepackage{orcidlink}
\usepackage{hyperref}

\newcommand{\iitj}{\affiliation{Department of Physics, Indian Institute of Technology Jodhpur\\
		N.H. 62, Nagaur Road, Karwar, Jodhpur, Rajasthan, India - 342030.}}

\begin{document}

\preprint{ }

\title{Anomalous Temperature Induced Transition and Convergence of Thermal Conductivity in Germanene Monolayer}

\author{Sapta Sindhu Paul Chowdhury$\,$\orcidlink{0009-0007-0472-7660}}
\iitj
\author{Sourav Thapliyal$\,$\orcidlink{0009-0003-3214-9184}}%
\iitj
\author{Santosh Mogurampelly$\,$\orcidlink{0000-0002-3145-4377}}%
\email{santosh@iitj.ac.in}
\iitj



\date{\today}

\begin{abstract}

We report an anomalous temperature-induced transition in thermal conductivity in germanene monolayer around a critical temperature $T_c = 350 \, \text{K}$. Equilibrium molecular dynamics simulations reveal a transition from $\kappa \sim T^{-2}$ scaling below $T_c$ to $\kappa \sim T^{-1/2}$ above, contrasting with conventional $\kappa \sim T^{-1}$ behavior. This anomalous scaling correlates with the long-scale characteristic timescale $\tau_2$ obtained from double exponential fitting of the heat current autocorrelation function. Phonon mode analysis using normal mode decomposition indicates that a redshift in ZO phonons reduces the acoustic-optical phonon gap, causing an overlap, enhances the phonon-phonon scattering, driving the anomalous scaling behavior. Moreover, nonequilibrium simulations find a convergent thermal conductivity of germanene with sample size, in agreement with mode coupling theory, owing to the high scattering of ZA phonons due to the inherent buckling of germanene.
\end{abstract}

\maketitle

\section{\label{sec:intro}Introduction\protect}
Germanene, the two-dimensional (2D) analog of germanium, is an experimentally realized distinctive 2D material with promising electronic and thermal properties \cite{Yuhara2020, Acun2015}. Composed of a single atomic layer of Ge atoms in a hexagonal lattice \cite{Roome2014}, germanene differs fundamentally from planar 2D materials like graphene, h-BN, etc., due to its intrinsic buckling, which introduces out-of-plane atomic displacements \cite{Cahangirov2009}. This inherent buckling not only enables tunable properties \cite{ARJMAND2018657} as a result of additional degrees of freedom but also enhances scattering of out-of-plane (flexural) phonons, a feature with significant implications in the thermal transport of germanene.

Several studies have focused on understanding the electronic properties of germanene \cite{Li2014_nano}. The electrical and thermal conductivity have been previously studied by Chegel et al. \cite{Chegel2020}. It is found that the band gap of germanene can be opened and increased to 20-310 meV \cite{Ye2014}.
The application of germanene is explored in the case of Li and Na ion batteries, resulting in high charge capacities \cite{Mortazavi2016}. Electronic properties due to external defects are found to be highly sensitive to vanishing Dirac cone \cite{Padilha2016}. Similarly, functionalization with H, I, F, Cl, and Br is proposed for the topological insulating devices based on germanene \cite{Si2014}.


The thermal transport of germanene is reported in a few works, compared to the extensive studies based on electronic properties. Previous work on the thermal properties of germanene reports a low thermal conductivity, attributed to its intrinsic buckling and relatively high atomic mass. Peng et al. \cite{Peng2016prb} reported a thermal conductivity of 2.4 W/(m K) at 300 K using the Boltzmann transport equation with \textit{ab-initio} force constants, while Das et al. \cite{das2019} found 3.1 W/(m K) using the Debye-Callaway model. These values contrast sharply with planar 2D materials, supporting the hypothesis that buckling reduces thermal transport efficiency \cite{Tang2021}. External strain, however, significantly enhances thermal conductivity, with pronounced size effects under increased strain \cite{Kuang2016}. Silicene-germanene superlattices also show a non-monotonic thermal conductivity due to phonon confinement effects \cite{WangNano2017}. Vibrational studies indicate that specific heat in germanene follows a $T^2$ dependency at low temperatures, further highlighting its unique thermal characteristics \cite{Zaveh2016}.

While the thermal transport properties of graphene have been extensively studied \cite{nika2017phonons, Balandin2008, lindsay2014phonon}, systematic investigations into buckled monolayer germanene are limited. An in-depth understanding of phonon propagation and thermal conductivity in such 2D materials is crucial for their optimization in high-performance applications. To date, the literature lacks comprehensive experimental data and theoretical analyses on thermal transport in germanene, particularly on the role of temperature-induced phonon relaxation effects. This study seeks to fill this gap, presenting detailed analyses of germanene’s temperature-dependent phonon thermal transport and its implications for future applications. 

This paper is organized as follows: Sec. \ref{sec:theory} provides an overview of the theoretical framework and computational methodology. In Sec. \ref{sec:results}, we present our findings: Sec. \ref{sec:structure} discusses the temperature-dependent structural behavior of germanene, followed by thermal conductivity calculations and their temperature dependence in Sec. \ref{sec:thermalcond}. Sec. \ref{sec:phonons} explores the mechanisms underlying phonon thermal transport, with a detailed analysis of phonon dispersion and linewidth, and their temperature dependence. Sec. \ref{sec:lengthdep} examines the influence of sample size on thermal conductivity. Finally, Sec. \ref{sec:conclusions} summarizes the key findings and their implications.

\section{\label{sec:theory}Computational Details\protect }
We use the Large-scale Atomic/Molecular Massively Parallel Simulator (Lammps) \cite{lammps}  for carrying out the molecular dynamics simulations. The Verlet algorithm is used for integrating Newton's equation of motion. A lot of available potentials for germanene monolayer, including the machine learning potentials, are compared in the literature for accuracy in classical simulations. Among the potentials studied, Tersoff, Modified Embedded Atom Method (MEAM), and Stillinger-Weber (SW) are recommended for simulating germanene \cite{Marcin2023}. However, Tersoff potential is unsuitable for our calculations as the systems simulated using this potential deform at 700 K \cite{Thapliyal2024}, which is inconsistent with the literature \cite{Peng2016prb}. So, the modified SW potential is used for modeling the interatomic interactions. We take the optimized parameters of SW potential for all the simulations from the work of Jiang et al \cite{Jiang2017}. The periodic boundary condition has been applied in all three directions. However, to make the out-of-plane direction of the germanene monolayer aperiodic, we have added a 50 {\AA} vacuum in the Z-direction. To obtain optimized geometry, molecular mechanics-based minimizations with the steepest descent and the conjugate gradient algorithms are carried out with a tolerance of 10$^{-14}$ for normalized energy difference and 10$^{-14}$ eV/{\AA} for force. To equilibrate the systems for thermal conductivity calculations, we have simulated the system in isothermal-isobaric (NPT) ensemble and thereafter in an isothermal (NVT) ensemble for 500 ps respectively (Fig. S1, S2 in SI) with 1 fs timestep for integrating Newton's equation of motion. We use the Nose-Hoover thermostat with a coupling constant of 0.1 ps and Nose-Hoover barostat with 1 ps coupling constant for maintaining temperature (100-950 K) and pressure (0 bar) throughout simulation runs. We calculate the thermal conductivity by ensemble averaging 30 independent trajectories, created by randomizing the velocity distribution of Ge atoms. Each trajectory is simulated for 2 ns in NVT ensemble (Fig. S3 in SI) with a fine timestep of 0.1 fs, which is more than sufficient to resolve the highest phonon frequency of the germanene monolayer \cite{Wang2015, Pei2013}.

In equilibrium formalism, the thermal conductivity ($\kappa$) at an ambient temperature \textit{T} is calculated by using the Green-Kubo relation by integrating the heat current auto-correlation function:

\begin{equation}
	\label{eqn:gk_cond}
	\bm{\kappa}(T) = \frac{1}{Vk_BT^2}\int_{0}^{\infty} \langle \bm{S}(0).\bm{S}(t) \rangle dt.
\end{equation}
Here, the volume is calculated as V=$L_x\times L_y\times$(4.24 \AA), where $L_x$ and $L_y$ are the dimensions of the simulation cell in the X and Y directions, and the van der Waals radius of Ge atoms (4.24 \AA) is taken as the thickness \cite{Mantina2009}. The temperature variation of thermal conductivity for the germanene monolayer is calculated using this method.
The heat current operator, S(t) is calculated by:

\begin{equation}
   S(t) = \sum_i E_iv_i+\sum_{i}\sigma_iv_i,
\end{equation}
with  $E_i$ is the energy, $v_i$ is the velocity of $i^{th}$ atom, and $\sigma_i$ represents the virial stress tensor calculated from both the two-body and three body term of the potentail \cite{Surblys2019, Surblys2021}. More details on the calculations of phonon density of state (PDOS) and dispersion relations are available in our previous works \cite{santosh2018,sapta2023}.

We use Fourier's law to calculate the thermal conductivity from the nonequilibrium simulations, given by:
\begin{equation}
	\label{eqn:fourier}
	J=-\kappa\nabla T,
\end{equation}
where J is the heatflux flowing from hotter to colder regions of the simulation.  \cite{Mantina2009}. The sample length variation of thermal conductivity for the germanene monolayer is calculated using this non-equilibrium-based method. We have used systems of length 10 nm to 10000 nm for the calculation of the length dependence of thermal conductivity. To achieve this, we have created 25$\times$25$\times$1 to  25000$\times$25$\times$1 supercells consisting of 1250 atoms to 1250000 atoms. The same equilibration protocol is followed as explained earlier. Unlike in the case of equilibrium simulations, for simulating the systems in non-equilibrium conditions, we fix one end of the sheet to a higher temperature of 320 K and the other end of the system at 280 K (Fig. S4) by using a Langevin thermostat, as implemented in LAMMPS. A total of 5 ns NVE run with 1 fs integration timestep is done; out of which the last 1.5 ns is used to calculate the average temperature and heatflux in a block of 10 {\AA} length.

The phonon power spectrum is calculated from the velocity auto-correlation function by \cite{dynaphopy}:
\begin{equation}
	G_{\bm{q}j}(\omega)= 2\int_{-\infty}^{\infty} \langle v_{\bm{q}j}^*(0)v_{\bm{q}j}(\tau)\rangle e^{i\omega\tau}d\tau,
\end{equation}
where the velocity autocorrelation function is:
\begin{equation}
	\langle v_{\bm{q}j}^*(0)v_{\bm{q}j}(\tau)\rangle = \lim_{t\prime\rightarrow\infty}\frac{1}{t\prime}\int_{0}^{t\prime}v_{\bm{q}j}^*(t+\tau)v_{\bm{q}j}(t)dt.
\end{equation}
Here, $v_{\bm{q}j}$ is the velocity of phonon branch $j$ corresponding to a wavevector $\bm{q}$, obtained from MD simulations. We use the normal mode decomposition technique to project atomic velocities on the wavevector $\bm{q}$:
\begin{equation}
	\bm{\textbf{v}}^{\bm{q}}_l(t)= \sqrt{\frac{m_l}{N}}\sum_{k}e^{-i\bm{q}.\bm{r}_{kl}^0}\bm{\textbf{v}}_{kl}(t).
\end{equation}
Projecting into phonon eigenvector $\textbf{e}(\textbf{q}, j)$, we have:
\begin{equation}
	v_{\bm{q}j}(t) = \sum_l\textbf{v}_l^{\bm{q}}(t).\textbf{e}_l^*(\bm{q},j).
\end{equation}
Now, the power spectrum is fitted with a Lorentzian to obtain the phonon linewidth $\gamma_{\bm{q}j}$ as:
\begin{equation}
	G_{\bm{q}j}\equiv\frac{\langle \left| v_{\bm{q}j}(t)\right|^2\rangle}{\frac{1}{2}\gamma_{\bm{q}j}\pi\left(1+\left(\frac{\omega-\widetilde{\omega}_{\bm{q}j}}{\frac{1}{2}\gamma_{\bm{q}j}}\right)^2\right)}.
\end{equation}

Here, $\omega$ is the frequency at peak, and $\gamma$ represents the half-width at half maximum. The phonon lifetime is calculated by $\tau = 1/(2\gamma)$.

\section{\label{sec:results}Results and Discussion\protect}
\subsection{\label{sec:structure}Effect of Temperature on Structural Geometry}

\begin{figure}[h]
	\includegraphics[height=6.5cm,keepaspectratio]{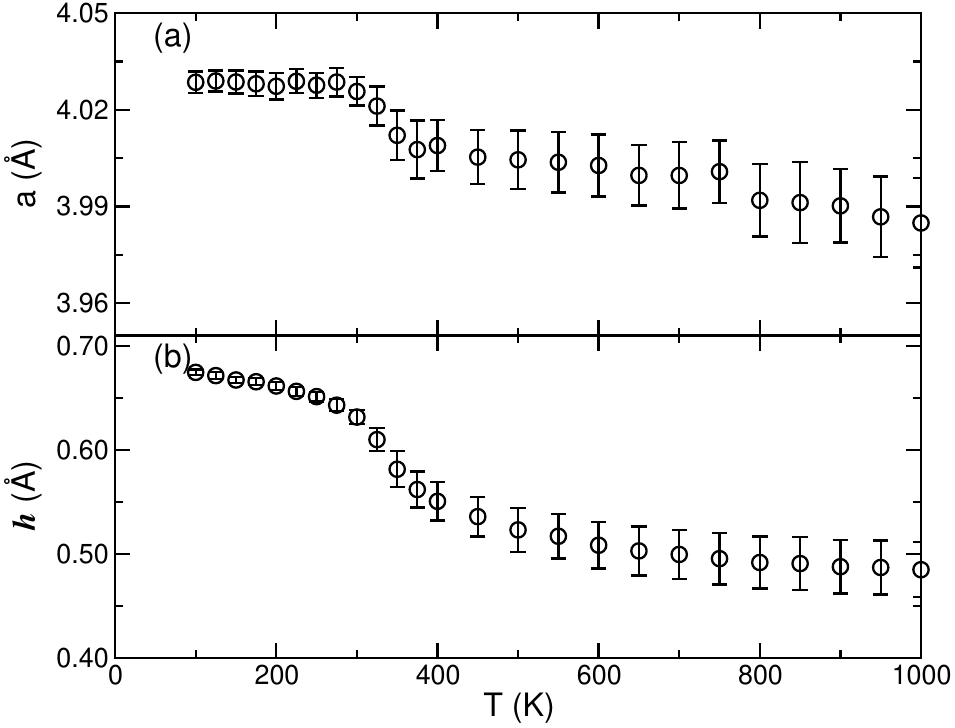}
	\caption{\label{fig:structure}(a) The lattice constant and (b) buckling height of a germanene monolayer as functions of temperature. The lattice constant and buckling height decrease monotonically, with a relatively rapid decline observed between 300 K and 400 K.}
\end{figure}

Germanene, comprising a hexagonal lattice of Ge atoms, exhibits intrinsic buckling that distinguishes it from planar 2D materials like graphene and h-BN. All structural parameters are based on the experimentally reported geometry \cite{expt-str, Yuhara2018}. Molecular mechanics optimization yields a lattice constant of $a =3.98$ \AA~and a buckling height of 0.71 \AA, with the system maintaining the symmetry operations of the P-3m1 space group. A 32$\times$32$\times$1 supercell containing 2048 Ge atoms is used for all simulations. This is sufficient for the calculation of thermal conductivity using MD simulations in buckled materials \cite{Li2012}.

To investigate the thermal effects on the lattice geometry, we conduct simulations at a fixed pressure of 0 bar in the NPT ensemble across a range of temperatures. Figure \ref{fig:structure} (a,b) illustrates the temperature dependence of the lattice constant and buckling height. We observe a monotonic decrease in both parameters with increasing temperature, though their rates of change vary across different temperature regimes. This atypical behaviour of monotonic decrease of both the lattice constant and the buckling height has been previously observed in plumbene, a group IV 2D material like germanene \cite{Mohammadi2024}. At lower temperatures, the lattice constant remains nearly constant, while the buckling height decreases. In the intermediate range (300-400 K), lattice constant and buckling height show a more rapid decline at comparable rates. Beyond 400 K, however, the lattice constant stabilizes, while the buckling height continues to decrease at a slightly faster rate.

This temperature-dependent behavior reflects a dual mechanism of structural relaxation. Initially, as thermal energy increases, germanene primarily relaxes through reductions in buckling height, which equilibrates internal stresses without significantly altering the in-plane lattice. Around 300 K, further temperature increases lead to a cooperative relaxation between lattice constant contraction and buckling height reduction. As temperature increases beyond 400 K, the competing relaxations stabilize, with the lattice constant decreasing at a slower rate and buckling remaining the dominant mode of structural adaptation.

\subsection{\label{sec:thermalcond}Anomalous Transition in Thermal Conductivity}

\begin{figure}[h]
	\includegraphics[height=6.5cm,keepaspectratio]{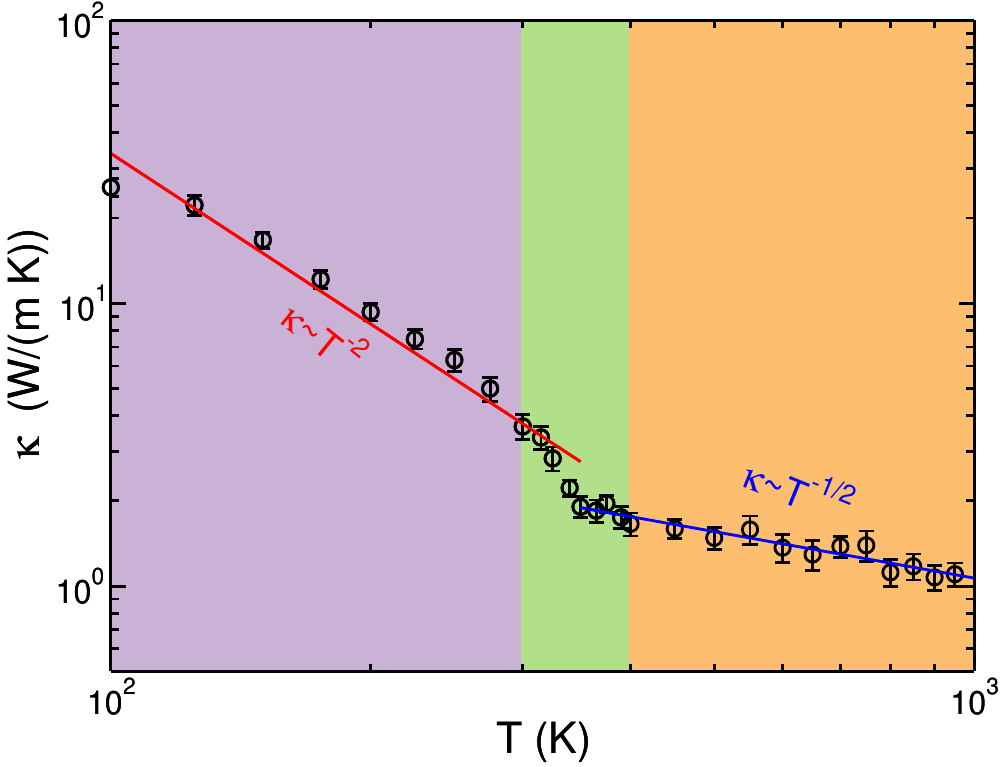} 
	\caption{\label{fig:kappavsT} Thermal conductivity of a germanene monolayer as a function of temperature. An anomalous transition in $\kappa$ is observed near $T_c\sim$ 350 K, with distinct power-law relations with exponents -2 and -1/2 below and above $T_c$, respectively.}
\end{figure}

The thermal conductivity of a germanene monolayer is determined using equilibrium molecular dynamics simulations (Eq. \ref{eqn:gk_cond}), yielding a value of $3.6 \pm 0.4 \, \text{W/m.K}$ at room temperature, in close agreement with prior studies \cite{Thapliyal2024,Peng2016prb, das2019}. As illustrated in Figure \ref{fig:kappavsT}, $\kappa$ demonstrates a monotonic decline with increasing temperature, typical of enhanced phonon-phonon scattering that limits heat transport at higher temperatures. However, a distinct and unexpected transition in thermal conductivity behavior occurs at a critical temperature $T_c \approx 350$ K, where the temperature scaling of $\kappa$ markedly shifts, revealing unique thermal transport characteristics in germanene not observed in most 2D materials. 

This transition is quantified by power-law fits to the data below and above $T_c$, showing that $\kappa \sim T^{-2}$ for temperatures below $T_c$ and $\kappa \sim T^{-1/2}$ for temperatures above $T_c$. Such behavior deviates from the conventional $\kappa \sim T^{-1}$ trend typical for crystalline 2D systems, \cite{Qin2017, Carruthers1961} where phonon scattering rates are expected to yield a single scaling regime. The observed $T^{-2}$ scaling below $T_c$ indicates an unusually rapid decay in thermal conductivity with temperature, suggesting that lower-temperature heat transport in germanene is highly sensitive to the interactions between specific phonon modes. In contrast, higher temperatures favor a reduced sensitivity, as evidenced by the more gradual $T^{-1/2}$ scaling above $T_c$.

It must be noted that no reliable quantum correction exists for classical thermal conductivity calculated from MD simulations \cite{Fan2017prb, Turney2017}. As a result, we do not consider the quantum corrections \cite{Turney2017} in this work. In graphene, thermal conductivity is underestimated by 10\% in classical calculations below the Debye temperature \cite{Fan2017prb, Singh2011}. But an increase in the thermal conductivity below 352 K (the Debye temperature of germanene) may increase the exponent $\alpha$ ($\sim$2 from the classical calculations) in the scaling relation $\kappa\sim\frac{1}{T^{\alpha}}$  below the transition temperature, whereas the exponent will remain the same in the higher temperature regime, which will still lead to the anomalous transition, as evident from our calculations.

The observed shift from $T^{-2}$ to $T^{-1/2}$ scaling at $T_c$ highlights a critical threshold in the thermal behavior of germanene, demarcating two distinct heat transport regimes. Below $T_c$, enhanced thermal resistivity dominates due to specific phonon scattering interactions, while above $T_c$, these processes diminish or reconfigure, leading to a more stable thermal regime. This anomalous transition is likely tied to the unique buckled structure of germanene, offering key insights into designing germanene-based devices. In the following, we explore these mechanisms by analyzing the relaxation phenomena and phonon dispersion, density of states, and linewidths.

\subsection{\label{sec:phonons}Mechanisms Underlying Phonon Thermal Transport}
\begin{figure}[h]
	\includegraphics[height=6.5cm,keepaspectratio]{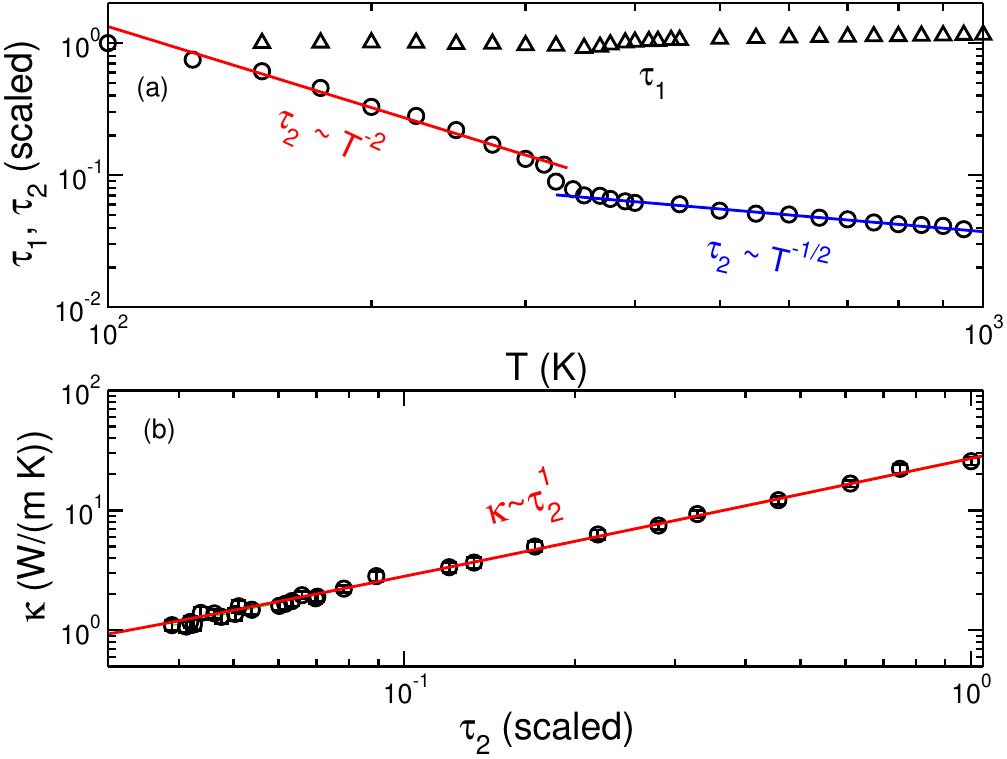}
	\caption{\label{fig:tauvsT} (a) Scaled long-scale (circles) and short-scale (triangles) characteristic times with temperature for germanene monolayer, derived from double-exponential HCACF fitting, and (b) thermal conductivity as a function of the long-scale characteristic time. The long-scale characteristic time follows the same temperature dependence as thermal conductivity, while the short-scale characteristic time dips near the transition temperature.}
\end{figure}
To investigate the underlying heat transport mechanisms in germanene, we analyze the nature of the heat current autocorrelation function (HCACF) through double-exponential fitting:
\begin{equation}
	\frac{\langle \bm{S}(0) \cdot \bm{S}(t) \rangle}{\langle \bm{S}(0) \cdot \bm{S}(0) \rangle} = \alpha_0 e^{-t/\tau_1} + (1 - \alpha_0) e^{-t/\tau_2},
\end{equation}
where $\tau_1$ is the short range characteristic timescale, $\tau_2$ is the long range characteristic timescale, and $\alpha_0$ (0 $\le \alpha_0 \le$ 1) is a fitting parameter. The short-range characteristic timescale ($\tau_1$) is typically attributed to the high-frequency (optical) phonons \cite{Grujicic2005}. The initial rapid decay in the HCACF is thus linked to the loss of correlation among these optical phonons. The long-range characteristic timescale ($\tau_2$) is associated with low-frequency (acoustic) phonons. The slow, long-time tail of the HCACF reflects the persistence of correlations among these acoustic phonons, which dominate the contribution to the total thermal conductivity \cite{Grujicic2005, Tretiakov2004, CHEN2010}

Fig. \ref{fig:tauvsT}(a) shows the temperature dependence of $\tau_1$ and $\tau_2$. The slow decay characteristic timescale $\tau_1$ decreases up to the transition temperature and then begins to increase, while $\tau_2$ decreases monotonically. While $\tau_1$ shows no clear monotonic trend with $T$, $\tau_2$ demonstrates a temperature-dependent pattern that is qualitatively analogous to the behavior of $\kappa$. A power-law fit yields $\tau_2 \sim T^{-2.04}$ at lower temperatures below $T_c$, closely mirroring the thermal conductivity behavior and $\tau_2 \sim T^{-0.57}$ at higher temperatures above $T_c$. In Figure \ref{fig:tauvsT}(b), thermal conductivity is plotted against $\tau_2$, showing a clear linear relationship that directly links the long-scale characteristic time to thermal transport. This strong correlation supports the central role of $\tau_2$ in governing the temperature-dependent thermal conductivity. While the double-exponential model elucidates a simplistic connection between thermal conductivity and HCACF decay, mode-dependent phonon analysis is required for a more comprehensive understanding.

\begin{figure}[h]
	\includegraphics[height=6.5cm,keepaspectratio]{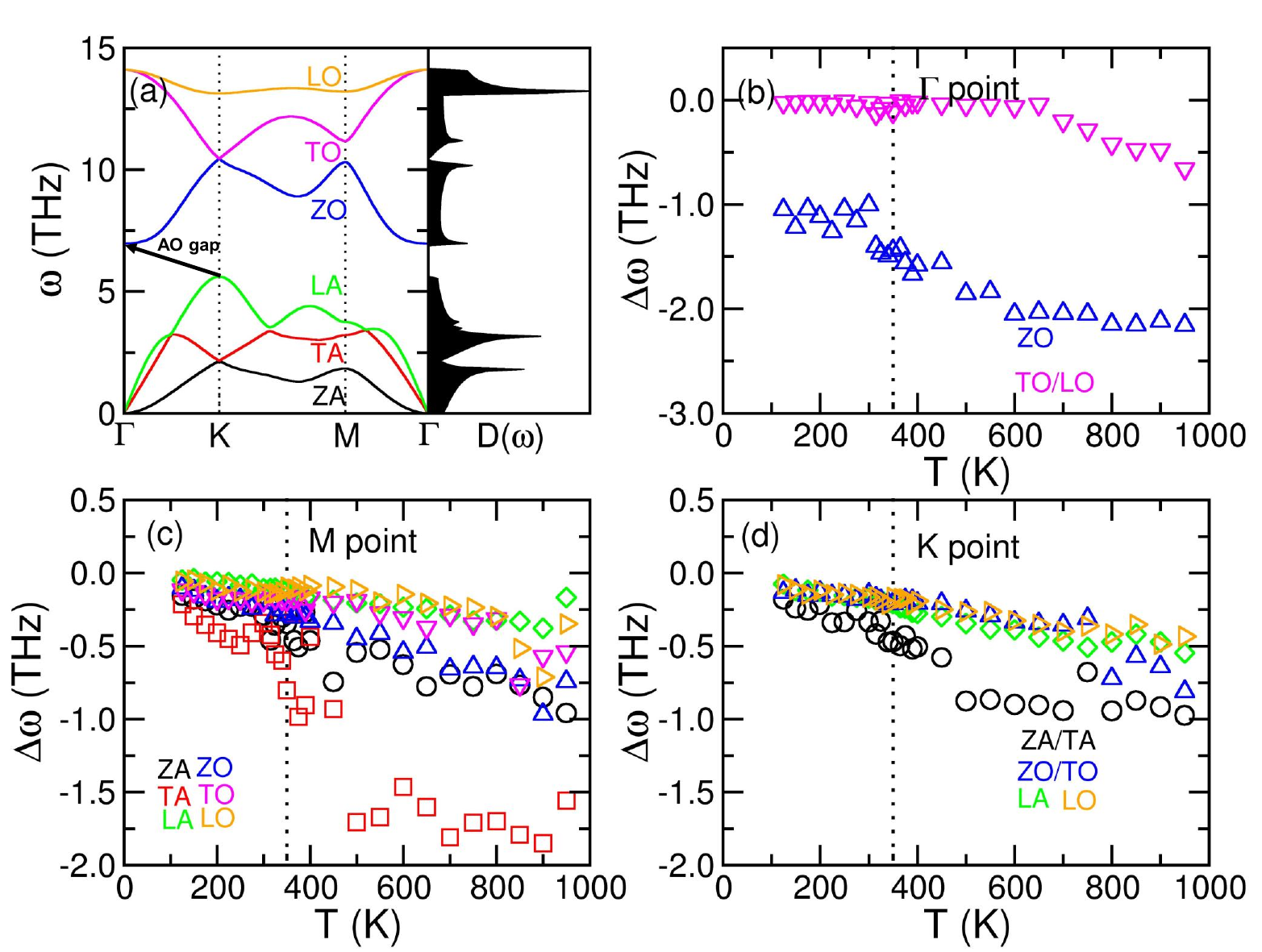}
	\caption{\label{fig:dispersion} (a) Phonon dispersion and density of states for germanene along $\Gamma\rightarrow \text{K} \rightarrow \text{M}\rightarrow\Gamma$ direction. The black arrow shows the acoustic-optical gap in the phonon dispersion. (b-d) Angular frequency deviation from harmonic approximation with temperature at high-symmetry $\Gamma$, M, and K points. Notably, the acoustic-optical gap narrows with temperature, contributing to thermal conductivity reduction. The dotted lines represent the transition temperature.}
\end{figure}

To explore further the thermal transport mechanisms in monolayer germanene, we analyze the phonon density of states (DOS) and dispersion along the $\Gamma \rightarrow \text{K} \rightarrow \text{M} \rightarrow \Gamma$ path (Fig. \ref{fig:dispersion}(a)). The maximum LO/TO frequency of 14.1 THz aligns with Jiang et al. \cite{Jiang2017}, and the absence of negative frequencies supports structural stability at simulated temperatures. As expected for 2D materials, the ZA mode exhibits quadratic dispersion, while the TA and LA modes show cubic behavior \cite{sapta_beryllene, Gu2015}. The DOS peak at 13.13 THz, dominated by LO phonons, undergoes a redshift with temperature, consistent with similar trends observed in other 2D materials \cite{santosh2018, sapta2023}.

To assess temperature effects on phonon dispersion, we plot frequency deviations at high-symmetry points in Fig. \ref{fig:dispersion}(b-d). In germanene, the acoustic-optical (AO) gap is the gap between the maximum phonon frequency of the LA mode at the K point and the optical ZO mode at the $\Gamma$-point, as evident from Fig. \ref{fig:dispersion}(a). At $\Gamma$ (Fig. \ref{fig:dispersion}(b)), acoustic modes remain unchanged due to symmetry, though the ZO mode shows significant redshift, which in turn reduces the AO gap to a point of overlap between the acoustic and optical phonon branches. Both TO and LO frequencies decrease modestly with temperature up to 650 K, after which redshifting intensifies. At the M point (Fig. \ref{fig:dispersion}(c)), the ZA and TA modes experience an abrupt decrease in frequency around the transition temperature (350 K), with the TA mode exhibiting particularly strong redshifting. While ZO modes show frequency reductions across temperatures, a minimal shift in LA modes is observed. At the K point, (Fig. \ref{fig:dispersion}(d)), the ZA modes show significant redshifting near the transition temperature, while the ZO/TO and LA modes are slightly decreased, leading to a decreased AO gap, and an overlap between the acoustic and optical branches at the transition point. In the case of the acoustic phonons, two scattering processes are applicable: (i) AAA, signifying the scattering between three acoustic phonons, and (ii) AAO, where two acoustic phonons are absorbed into one optical phonon and vice versa. A large AO gap weakens the AAO scattering channel, and in some cases, they might be prohibited \cite{Lindsay2013}. As a result, a reduction in AO gap due to the increase in temperature leads to a stronger AAO scattering channel and a decreased thermal conductivity \cite{Lindsay2015}. Overall, the AO gap narrowing with temperature significantly impacts thermal conductivity, similar to the observations in other 2D materials \cite{Lindsay2013, Gu2013}. The sharp AO gap reduction resulting from the significant redshifting of the optical ZO branch at the $\Gamma$-point likely correlates with the anomalous thermal conductivity transition. Moreover, from the phonon density of states calculation (Fig. S5 in SI), it is evident that the AO gap decreases with temperature. An overlap between the acoustic and optical phonon occurs when the temperature is raised to 350 K. The overlap increases with a further increase in the temperature, correlating with the transition.

\begin{figure}[t]
	\includegraphics[height=6.5cm,keepaspectratio]{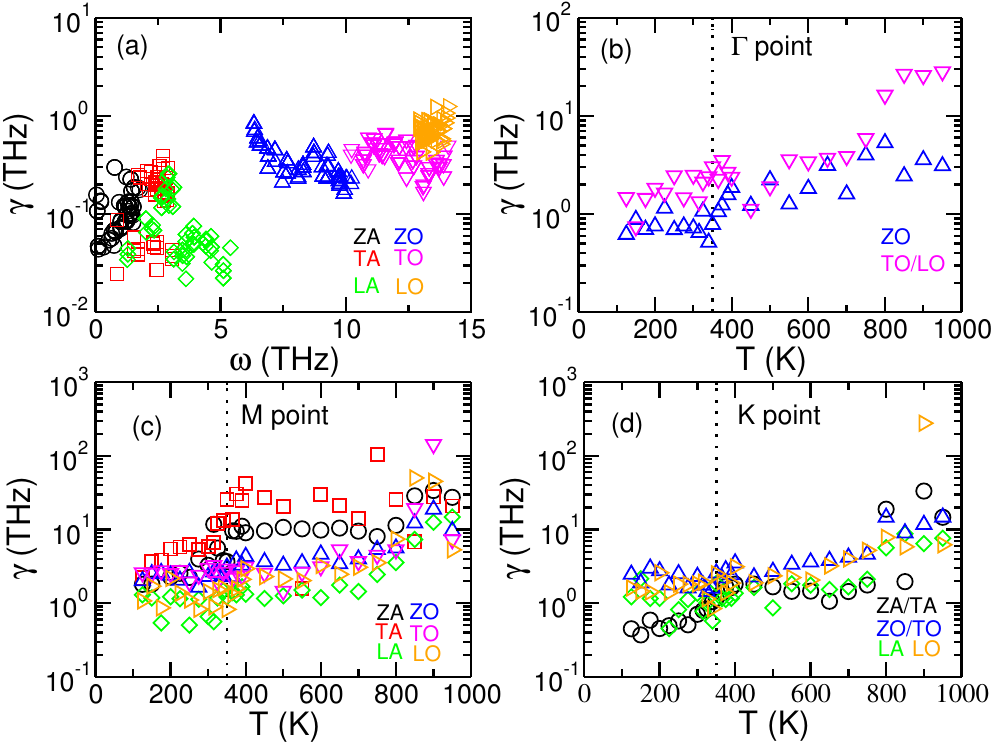}
	\caption{\label{fig:linewidth} (a) Mode-specific phonon linewidth of germanene at room temperature as a function of angular frequency. Phonon linewidth variation with temperature at (b) $\Gamma$, (c) M, and (d) K points. Linewidth increases with temperature, reducing phonon lifetimes and thermal conductivity. The dotted lines represent the transition temperature.}
\end{figure}
In Fig. \ref{fig:linewidth}(a), we present the linewidth of different phonon branches with phonon angular frequency at room temperature. The acoustic branches, primarily responsible for thermal transport, are found to possess lower linewidths than optical modes. Intriguingly, unlike planar 2D materials like graphene, where the ZA mode displays suppressed linewidth, the out-of-plane and in-plane acoustic modes in germanene exhibit comparable linewidths at 300 K. The temperature-dependent phonon linewidths are plotted for $\Gamma$, M, and K points in Fig. \ref{fig:linewidth}(b-d). Moreover, we plot long-wavelength acoustic phonons for a few critical temperatures in Fig. S6 in the SI. Further, in Fig. S7, S8, S9, we plot the phonon linewidth of different phonon branches at the $\Gamma$, M, and K points separately for clarity. We find that the phonon linewidths increase with temperature, reflecting intensified phonon scattering at elevated temperatures. At $\Gamma$-point, optical phonon linewidths, particularly ZO, change significantly near 350 K (Fig. \ref{fig:linewidth}(b)). The acoustic modes show a significant increase in linewidth at the M point. From Fig. \ref{fig:linewidth}(c), it is evident that the linewidth of ZA and TA increases significantly at this symmetry point. Moreover, a clear transition is observed for TA and ZA phonons near the transition temperature, as evident from Fig. S10. This increase in phonon linewidth reduces the phonon lifetime, thus reducing the thermal conductivity significantly. Similarly, at the K point, the ZA/TA acoustic modes show a significant deviation (Fig. S11) in linewidth near 350 K (Fig. \ref{fig:linewidth}(d)). This pronounced followed by a steady increase of acoustic phonon linewidths underpin the observed thermal conductivity transition. In the transition regime, both the lattice constant and buckling decrease sharply, which correlates with a direct decrease in the thermal conductivity. A decrease in the buckling height actually increases the contribution of ZA phonons while reducing the contribution from high-frequency LA phonons \cite{Peng2016prb}. Typically, the flattening of buckling increases the thermal conductivity, whereas the decrease in the lattice constant leads to a decrease in the thermal conductivity \cite{Peng2016prb, Keshri2021}. Based on the variation of thermal conductivity with temperature, it is likely that among these contradictory effects, the decrease in the conductivity outplays the effect of a possible increase in the conductivity due to reduced buckling.

\subsection{\label{sec:lengthdep}Effect of Sample Size on $\kappa$}
\begin{figure}[h]
	\includegraphics[height=6.5cm,keepaspectratio]{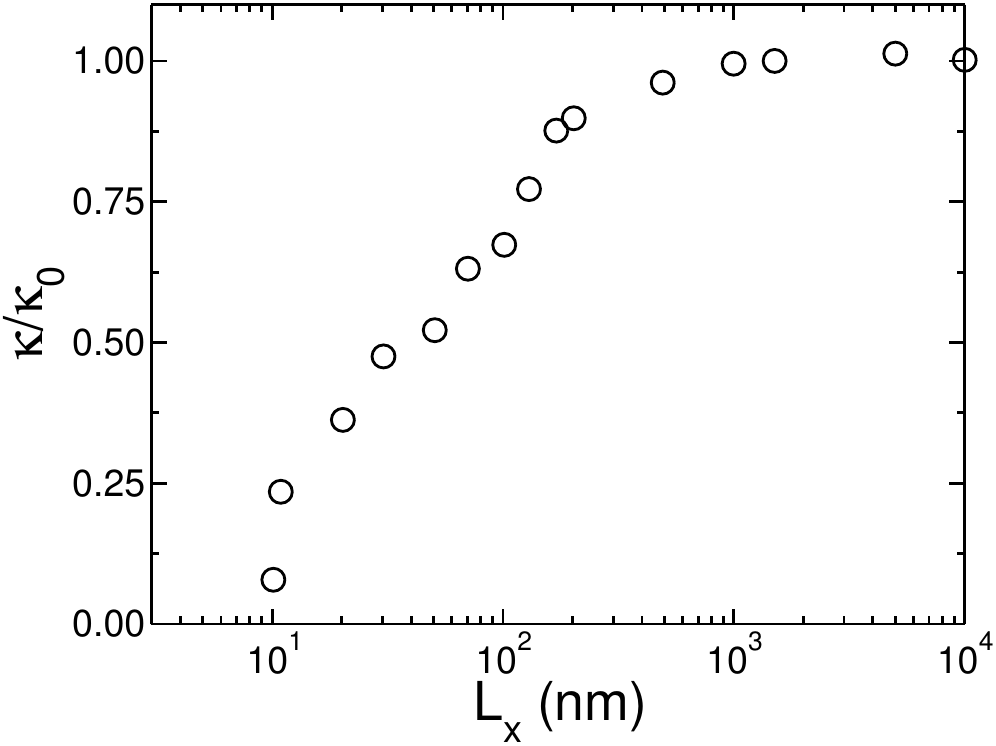}
	\caption{\label{fig:kappavsL} Thermal conductivity of a germanene monolayer as a function of sample length, calculated using NEMD simulations at room temperature and scaled to thermal conductivity at a 1 $\mu$m sample length. A convergence in thermal conductivity is observed due to pronounced out-of-plane scattering.}
\end{figure}

In low-dimensional materials, the thermal conductivity often exhibits a pronounced dependence on the sample dimensions, challenging the assumption of intrinsic behavior. 
In such cases, it is important to investigate the effect of sample length on the thermal transport in these materials. Nonequilibrium-based approach (Eq. \ref{eqn:fourier}) is suitable for understanding the length dependence of thermal conductivity as the approach can simulate a microdevice measurement on a sample placed between hot and cold regions (Fig. S4(a) in SI) \cite{Gu2018}.

In Fig. \ref{fig:kappavsL}, we present the thermal conductivity of monolayer germanene as a function of sample length, calculated at 300 K. Thermal conductivity increases consistently with length up to 500 nm before saturating at approximately 1 $\mu$m. This is in agreement with the thermal conductivity of germanene nanoribbons, calculated previously (Fig. S4(b)) \cite{Peng2016prb}. According to mode coupling theory, the behavior of the HCACF can indicate convergence or divergence in thermal conductivity, with $t^{-1}$ scaling signifying divergence and $t^{-1.5}$ or faster indicating convergence \cite{Lepri2003}. Here, the scaled HCACF exhibits a decay of $t^{-1.6}$ (Fig. S12 in SI), indicating convergent thermal conductivity.

This convergence is attributed to the reduced contribution from flexural phonons, influenced by germanene's inherent buckling. Fig. \ref{fig:linewidth}(a) shows that the out-of-plane and in-plane acoustic phonon branches exhibit similar linewidths due to symmetry breaking caused by buckling. Consequently, the ZA phonons with similar lifetime (shown in Fig. S13(a)) in SI contribute less to the overall thermal conductivity (only $\sim$33\% to total conductivity, compared to $\sim$ 76\% in graphene \cite{Peng2016prb, Fan2017prb}), leading to convergence at larger sample sizes. Further, large wavelength ZA phonons show $\omega \sim q^{2}$ scaling (Fig. S13(b) in SI), which is consistent with convergent thermal conductivity \cite{Gu2018}.

\section{\label{sec:conclusions}Conclusions\protect}
In conclusion, we have systematically investigated the thermal conductivity mechanism of monolayer germanene using molecular dynamics simulations within the Green-Kubo formalism across an extensive temperature range, employing an optimized Stillinger-Weber potential parameter set. We report an anomalous temperature-induced transition in the thermal conductivity of germanene monolayer at a critical temperature of 350 K. Our analysis reveals that the thermal conductivity follows an unusual scaling: $\kappa \sim T^{-2}$ below $T_c$, transitioning to $\kappa \sim T^{-1/2}$ above $T_c$, a stark departure from the typical $\kappa \sim T^{-1}$ trend. This anomalous scaling is closely mirrored in the slow relaxation timescale $\tau_2$ of the HCACF, with $\tau_2 \sim T^{-2}$ below $T_c$ and $\tau_2 \sim T^{-1/2}$ above, confirming the direct relationship $\kappa \sim \tau_2$.

Phonon analysis further elucidates this behavior, with a significant reduction in the AO gap, primarily due to redshifting of the ZO phonon modes at high-symmetry $\Gamma$ point. This redshifting, coupled with a significant deviation in acoustic phonon linewidths near $T_c$, signifies intensified phonon scattering, fundamentally altering the thermal transport pathways. Additionally, NEMD simulations indicate that the thermal conductivity of germanene converges with sample length, a characteristic attributed to comparable contributions from both in-plane and out-of-plane phonon modes, consistent with mode-coupling theory.

\begin{acknowledgments}
The authors acknowledge the Computer Center of IIT Jodhpur for providing computing resources that contributed to the research results reported in this paper.
\end{acknowledgments}

\bibliography{germanene_monolayer_manuscript}
\end{document}